\documentclass[preprintnumbers,showpacs,floats,twocolumn,prd,aps]{revtex4}
\usepackage{amssymb,amsmath}
\usepackage{epsfig,hyperref}

\makeatletter
\renewcommand{\vec}[1]{\mbox{\boldmath$#1$}}
\newcommand{\arXiv}[2][]{\href{http://arxiv.org/abs/#2}{\texttt{arXiv:#2\@ifempty{#1}{}{ [#1]}}}}
\makeatother

\begin{document}

\title{A Singularity Problem with f(R) Dark Energy}
\author{Andrei V. Frolov}\email{frolov@sfu.ca}
\affiliation{
  Department of Physics,
  Simon Fraser University\\
  8888 University Drive,
  Burnaby, BC Canada
  V5A 1S6
}
\date{March 13, 2008}

\begin{abstract}
  In this paper, I point out that there is a curvature singularity problem
  appearing on non-linear level that generally plagues $f(R)$ models that
  modify Einstein gravity in the infrared. It is caused by the fact that for
  the effective scalar degree of freedom, the curvature singularity is at a
  finite field value and energy level, and can be easily accessed by the field
  dynamics in the presence of matter. This problem is invisible in linearized
  analysis, except for the tell-tale growing oscillatory modes it causes. In
  view of this, viability of many $f(R)$ models in current literature will have
  to be re-evaluated.
\end{abstract}

\preprint{SCG-2008-01}
\maketitle

What is causing the observed accelerated expansion of the Universe today is
one of the biggest open question in modern cosmology. Trying to explain it by
modifying theory of gravity rather than by introducing a mysterious dark
energy component has been a popular pursuit as of late. Unfortunately, it is
proving to be a rather difficult thing to do consistently, while avoiding
variety of stringent observational tests of gravity we have at our disposal.
A class of such models that received much attention recently is the one
which modifies Einstein-Hilbert gravitational action by replacing Ricci
curvature scalar by an arbitrary function of the curvature
\begin{equation}\label{eq:action}
  S = \int \left\{ \frac{f(R)}{16\pi G} + {\cal L}_{\text{m}} \right\} \sqrt{-g}\, d^4x.
\end{equation}
Introduced in cosmological context for the case which modifies gravity in the
high energy limit in a seminal paper by Starobinsky~\cite{Starobinsky:1980te}
and studied in \cite{Kofman:1987ec,Whitt:1984pd,Maeda:1988ab}, this model has later been
adopted for infrared modifications of gravity as well \cite{Capozziello:2003tk,
Carroll:2003wy}. For the latter application, it turned out to be not without
problems. Certain constraints have to be imposed on function $f(R)$ for the
model to be linearly stable \cite{Dolgov:2003px} and cosmologically viable
\cite{Chiba:2006jp, Amendola:2006we, Sawicki:2007tf}. The first attempts
failed these constraints right away, but since then, models that evade them
have been found (for example see \cite{Hu:2007nk, Starobinsky:2007hu,
Appleby:2007vb} and references therein) and enough trust has been placed in
their viability to study cosmological structure formation in detail
\cite{Bean:2006up, Pogosian:2007sw}.

In this paper, I point out a serious curvature singularity problem that
affects many, if not all, infrared-modified $f(R)$ models. Being non-linear in
nature, it escaped scrutiny so far.

As it is well known, a new scalar degree of freedom appears in $f(R)$ gravity
that is not there in Einstein theory (sometimes dubbed the {\em scalaron}).
Conformal transformation of the metric can be employed to make it explicit in
the action \cite{Whitt:1984pd, Maeda:1988ab}. In this paper, I will avoid doing that to
keep the usual matter coupling to the metric, and work with the action
(\ref{eq:action}) directly. Variation with respect to metric yields
gravitational equations of motion
\begin{equation}\label{eq:eom}
  f' R_{\mu\nu} - f'_{;\mu\nu} + \left(\Box f' - \frac{1}{2}\, f\right) g_{\mu\nu} = 8\pi G\, T_{\mu\nu},
\end{equation}
where prime ($'$) denotes the derivative of the function $f$ with respect to
its argument $R$, and $\Box$ is the usual notation for covariant D'Alembert
operator $\Box \equiv \nabla_\alpha \nabla^\alpha$. The equation of motion for
a new scalar degree of freedom is given by the trace of equation
(\ref{eq:eom})
\begin{equation}\label{eq:trace}
  \Box f' = \frac{1}{3}\, (2f-f'R) + \frac{8\pi G}{3}\, T.
\end{equation}
Identifying the scalar degree of freedom explicitly by a variable redefinition
\begin{equation}\label{eq:phi}
  \phi \equiv f'-1,
\end{equation}
the equation (\ref{eq:trace}) above is cast in the form of equation of motion
of a canonical dimensionless scalar field $\phi$ with a potential $V$ and a
force term ${\cal F}$
\begin{equation}
  \Box\phi = V'(\phi) - {\cal F}.
\end{equation}
The effective scalar field potential $V(\phi)$ is determined by
\begin{equation}
  V'(\phi) \equiv \frac{dV}{d\phi} = \frac{1}{3}\, (2f-f'R)
\end{equation}
expressed in terms of the scalar variable $\phi$. In practice, given $f(R)$,
it is usually difficult to invert the definition of the scalar degree of
freedom (\ref{eq:phi}) explicitly, so it might be more convenient to
determine effective potential $V$ in a parametric form instead. By integrating
\begin{equation}\label{eq:pot}
  \frac{dV}{dR} \equiv \frac{dV}{d\phi} \frac{d\phi}{dR} = \frac{1}{3}\, (2f-f'R) f'',
\end{equation}
potential $V(\phi)$ is then given by a pair of functions
$\{\phi(R),V(R)\}$. The force term ${\cal F}$ that drives the scalar field
$\phi$ is a trace of the stress-energy tensor $T$, which for perfect fluid is
simply
\begin{equation}
  {\cal F} = \frac{8\pi G}{3}\, (\rho - 3p).
\end{equation}

Let us consider a homogeneous cosmological model in $f(R)$ gravity, with the
usual complement of matter fields. Expansion of the Universe is described by a
flat Friedman-Robertson-Walker metric
\begin{equation}\label{eq:frw}
  ds^2 = -dt^2 + a^2(t)\, d\vec{x}^2,
\end{equation}
and the scalar gravitational degree of freedom $\phi$ obeys a usual scalar
field equation, albeit with a force term on the right hand side
\begin{equation}\label{eq:box}
  \ddot{\phi} + 3H\dot{\phi} + V'(\phi) = {\cal F}.
\end{equation}
The analog of Friedman equation in $f(R)$ cosmology is not so transparent. Let
us consider $tt$ component of gravitational equations of motion
(\ref{eq:eom}). For metric (\ref{eq:frw}), it is
\begin{equation}\label{eq:tt}
  3 H (f')\dot{~} - 3\, \frac{\ddot{a}}{a}\, f' + \frac{1}{2}\, f = 8\pi G\, \rho.
\end{equation}
Note that unlike the usual Friedman equation, higher derivatives of scale
factor $a$ appear. Second derivative $\ddot{a}$ is written out explicitly, and
a third derivative is hiding in a time derivative of $f'$ term, which itself
contains Ricci curvature, and hence $\ddot{a}$. Seeing a second derivative of
the scale factor, one might be tempted to treat the above equation
(\ref{eq:tt}) as a dynamical evolution equation for the scale factor. Doing
so, however, is not a very good idea. For small deviations from Einstein
gravity, the coefficient in front of $\ddot{a}$ goes degenerate, and the
equation (\ref{eq:tt}) does not have a good limit determining $\ddot{a}$
(which is not all that surprising, considering that Friedman equation in
Einstein gravity does not constrain $\ddot{a}$ directly). To get a proper
limit, let us instead get rid of $\ddot{a}$ in favour of the curvature scalar
\begin{equation}
  R = 6 \left( \frac{\ddot{a}}{a} + \frac{\dot{a}^2}{a^2} \right).
\end{equation}
After that is done, the equation (\ref{eq:tt}) becomes
\begin{equation}\label{eq:H}
  H^2 + (\ln f')\dot{~} H + \frac{1}{6} \frac{f-f'R}{f'} = \frac{8\pi G}{3f'}\, \rho,
\end{equation}
and its role as a constraint equation is revealed. In the limit of Einstein
gravity $f' \rightarrow 1$, and so the last two terms on the left hand side
disappear, and one is left with the usual Friedman equation. In the general
case, the extra terms are functions of scalar degree of freedom $\phi$ and its
first time derivative. No higher derivatives appear in this equation anymore.

Thus the following simple picture of dynamics in the $f(R)$ cosmology emerges.
Above the infrared modification scale $R_0$, the expansion rate of the
Universe is set primarily by the matter density, just like in the usual
cosmology, with small corrections. Only once the local curvature drops below
$R_0$, the expansion rate starts feeling the effect of gravity modification.
The spacetime curvature, on the other hand, is controlled by the scalar degree
of freedom $\phi$ which gravity acquires. It obeys the usual scalar field
equation (\ref{eq:box}) with potential $V(\phi)$, the shape of which is
directly determined by function $f(R)$, and a driving term from the trace of
matter stress-energy tensor.

But here is the problem: it turns out that precisely those functions $f(R)$
that lead to Einstein-like gravity action in the large curvature regime, yield
a potential $V$ with an unprotected curvature singularity.

\begin{figure}
  \centerline{\epsfig{file=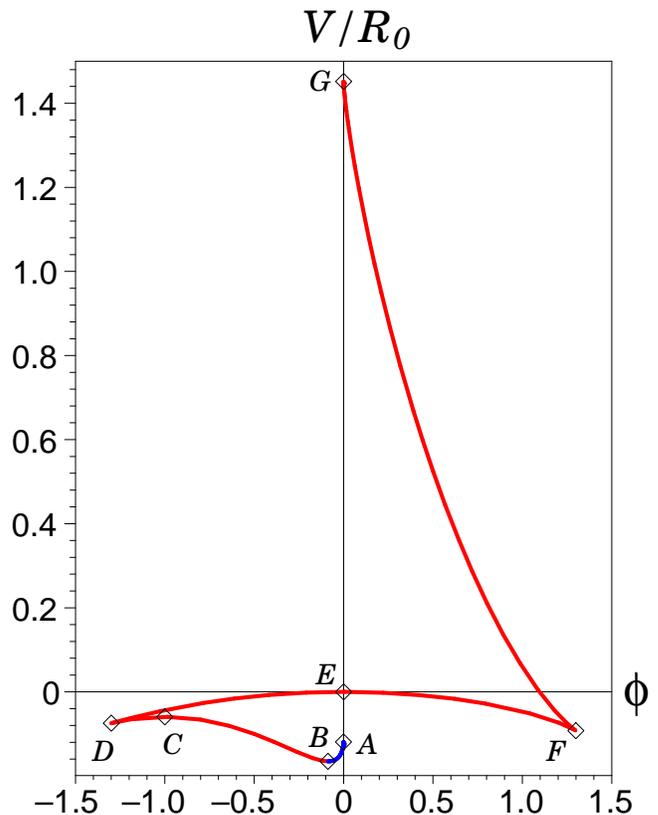, width=246pt}}
  \caption{
    Effective potential of a scalar degree of freedom in $f(R)$ gravity model
    (\ref{eq:star}) with $\lambda=2$ and $n=1$. Diamonds mark the location of
    critical points. The part relevant to cosmological evolution is emphasized
    by thick blue line.
  }
  \label{fig:pot}
\end{figure}

As a case in point, consider Starobinsky's disappearing cosmological constant
model \cite{Starobinsky:2007hu}, which has been very carefully constructed,
and avoids all known linear instabilities. It is described by
\begin{equation}\label{eq:star}
  f(R) = R + \lambda\left[\left(1+R^2\right)^{-n} - 1\right],
\end{equation}
where I have taken a liberty to absorb the cross-over curvature scale $R_0$
into rescaling of coordinates (which become dimensionless and are measured in
length units corresponding to $R_0$). For definiteness, let us take $n=1$. The
scalar degree of freedom in this model is given by
\begin{equation}
  \phi = - \frac{2\lambda R}{(1+R^2)^2}
\end{equation}
in terms of curvature, so large curvature limit $R \rightarrow \pm \infty$
corresponds to $\phi = 0$. Flat spacetime with $R=0$ also corresponds to
$\phi=0$, which gives us a hint that the potential is going to be a
multi-valued function. The potential can be evaluated by integrating
(\ref{eq:pot}); up to an arbitrary constant it is
\begin{eqnarray}
  V &=& \frac{\lambda^2 R (3+11 R^2+21 R^4-3R^6)}{24 (1+R^2)^4} \\
    &-& \frac{\lambda R^2 (1+R^2-R^4-R^6)}{3 (1+R^2)^4} - \frac{\lambda^2}{8} \arctan R. \nonumber
\end{eqnarray}
The effective scalar potential is plotted in Figure~\ref{fig:pot} for
$\lambda=2$, and is indeed multivalued. Let us walk through the interesting
locations on this plot. Point $A$ is a positive curvature singularity
$R=+\infty$. Point $B$ is the stable de Sitter minimum in this model, and
point $C$ is the unstable de Sitter maximum; their curvatures depend on
$\lambda$. Point $E$ corresponds to a flat spacetime, which although a
solution in this model, is unstable. Points $D$ and $F$ are critical points
with $f''=0$ that occur at $R=\pm 1/\sqrt{3}$; potential branches there.
Finally, point $G$ is a negative curvature singularity $R=-\infty$. Only the
small part of this potential is actually relevant for cosmological evolution
from initial singularity to today, and it lies in the arc $AB$, shaded blue in
the Figure~\ref{fig:pot}.

\begin{figure}
  \centerline{\epsfig{file=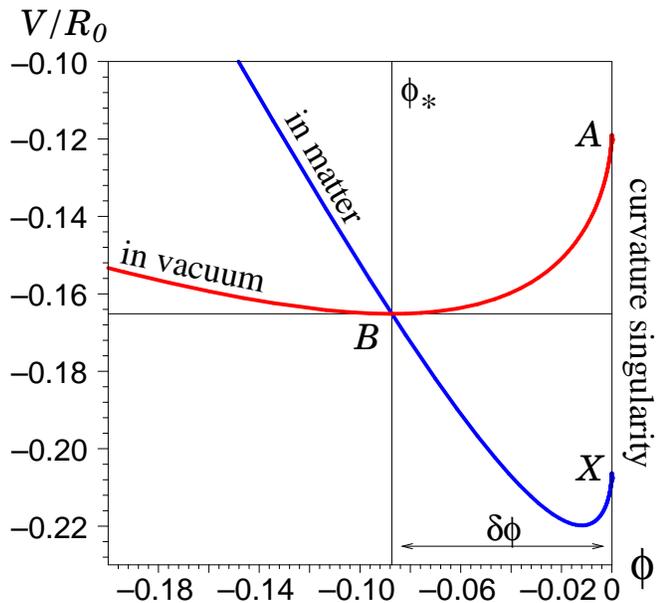, width=246pt}}
  \caption{
    Adding matter destabilizes the vacuum. Although effective potential inside
    constant density matter distribution still has a minimum, it is very
    shallow, and cannot protect the field $\phi$ from reaching curvature
    singularity $X$, which becomes energetically accessible from asymptotic
    vacuum state $B$.
  }
  \label{fig:shift}
\end{figure}

The most striking feature of the potential in Figure~\ref{fig:pot}, and the
core of the problem for infrared-modified $f(R)$ models, is that curvature
singularity at point $A$ is finite distance away both in field and energy
values from the place we are supposed to live in. Scalar degree of freedom
$\phi$ directly feels the matter distribution through the force term; for
equation of state $w<1/3$ the force is directed to the right, and drives the
field $\phi$ up the wall toward point $A$ and infinite curvature.
Characteristic scale of the potential $V$ is the cross-over curvature scale
$R_0$, and hence of the same order of magnitude as a present day cosmological
constant, which is exceedingly low compared to matter densities we encounter
every day. Given the scales involved, it appears to be quite easy to
over-drive the scalar degree of freedom and make it ``jump out'' of the
potential well by doing simple manipulations with normal matter (say a pile of
dust), which would cause catastrophic curvature singularity. Needless to say,
if this were to happen, it would not make for a desirable (or even viable)
model. Similarly, but less dramatically, matter with sufficiently stiff
equation of state can destabilize the model by driving the field to the left
past the unstable point $C$.

The presence of the curvature singularity a finite distance away is extremely
disturbing by itself, but let us examine more carefully if it is reached by
physically reasonable solutions. Inside a constant density matter
distribution, one can think of a (constant) force term ${\cal F}$ as coming
from a linear field potential ${\cal F}(\phi_*-\phi)$ instead, and introduce a
new ``in matter'' effective potential
\begin{equation}
  U(\phi) = V(\phi) + {\cal F}(\phi_*-\phi),
\end{equation}
where $\phi_*$ denotes the asymptotic de Sitter vacuum field. The comparison
between the two potentials $V$ and $U$ for ${\cal F} = R_0$ is shown in the
Figure~\ref{fig:shift}. As you can see, addition of matter slopes the
potential $U$, shifts the (stable) minimum to the right but makes it more
shallow, and lowers the curvature singularity point $X$. The density needed to
make the curvature singularity energetically accessible from vacuum $B$ is
given by the ratio of potential barrier $\delta V = V_A-V_B$ to the scalar
field value distance from vacuum to singularity $\delta\phi =
\phi_A-\phi_B=-\phi_*$. It is of the same order as the density of dark energy
today, with a numerical factor which depends on the model. So for vast
majority of physical solutions with matter, the curvature singularity is
energetically accessible from asymptotic de Sitter vacuum, and the potential
minimum is so close to curvature singularity that it would be invisible if
actually plotted to scale.

Energetical accessibility of the curvature singularity causes problems. For
example, if one takes a cosmological solution approaching dark energy
domination today, and traces it back into the past, one is very likely to
encounter a curvature singularity. This has been noticed numerically
\cite{Tsujikawa:2007xu, Appleby:2008tv}, but the underlying reasons for it and the extent of the
damage were not fully realized. It is also most likely the cause of growing
oscillatory curvature modes \cite{Starobinsky:2007hu} which signal the
break-down of linear expansion due to closeness of potential minimum to
curvature singularity. Although a more detailed analysis of the approach to
singularity is in order, from equation (\ref{eq:H}) it appears that the
singularity occurs at finite redshift, density and expansion rate, and is
driven by the divergence of the second derivative of the scale factor
$\ddot{a}$ (which would make it rather weak, but a singularity nonetheless).

Although I focused on cosmology so far, perhaps a more deadly argument
against having a curvature singularity at finite distance in the field space
comes from considering a gravitational field of a static dense compact object
(like a neutron star). Although the exact non-linear solution
of this problem is more complicated to analyze \cite{Kainulainen:2007bt,
Kainulainen:2008pr} and is beyond the scope of this article, I can give a
very simple estimate if the problem occurs. As we have seen, the energetics of
the scalar gravitational degree of freedom are by far dominated by the matter
driving term. If we discard the contribution of non-linear potential $V$
(which is negligible everywhere except maybe very close to singularity at
$\phi=0$ for compact object), the equation for gravitational field of a static matter
distributions becomes a simple Laplace equation
\begin{equation}
  \Delta\phi = - \frac{8\pi}{3} G \rho.
\end{equation}
Comparing this with an equation for Newtonian gravitational potential
\begin{equation}
  \Delta\Phi = 4\pi G \rho,
\end{equation}
we get a simple estimate for the excitation of scalar gravitational degree of freedom in
$f(R)$ gravity in terms of Newtonian potential well depth $\Phi$ of the compact object
\begin{equation}
  \phi = \phi_* - \frac{2}{3}\, \Phi,
\end{equation}
where $\phi_*$ is the asymptotic value of $\phi$ at infinity, i.e.\ the
minimum value $\phi_B = -\delta\phi$. But unlike Newtonian potential $\Phi$,
which has to diverge to cause singularity, or reach $-1/2$ to form a horizon,
gravitational degree of freedom $\phi$ needs only to change by a (small)
amount $\delta\phi$ from its vacuum value to create a singularity. So unless
an infrared-modified $f(R)$ model leads to a potential with curvature
singularity separated from vacuum by at least
\begin{equation}\label{eq:bound}
  \delta\phi \gtrsim \frac{1}{3},
\end{equation}
one would end up with a curvature singularity {\em without horizon} in a compact
astrophysical object like a neutron star. This condition is rather easy to
violate unless special care is taken in model-building. For example, for
Starobinsky's model (\ref{eq:star}) with $n=1$ and $\lambda=2$ (as in
Figure~\ref{fig:pot}) $\delta\phi \simeq 0.0874 \ll 1/3$, and is even smaller
for larger values of $\lambda$, for which it decreases as $\delta\phi \sim
(2\lambda)^{-2}$. Since in general one needs $f'>0$ for graviton not to be a
ghost, one would need $-1<\phi_* \lesssim -1/3$ to avoid both problems, the
prospects of achieving which without fine-tuning do not look good.

This curvature singularity problem is in no way unique to Starobinsky's
disappearing cosmological constant model \cite{Starobinsky:2007hu}, which I
have taken as an example simply because it is one of the most carefully
constructed models so far. In fact, any infrared-modified $f(R)$ gravity model
suffers from it. Let us consider arbitrary function $f(R)$, and require that
it reduces to Einstein gravity for large curvature, and has an analytic
expansion
\begin{equation}
  f(R) = R + \Lambda + \frac{1}{R^\alpha} \sum\limits_{n=0}^{\infty} \frac{\mu_n}{R^n}
\end{equation}
with a leading term $\mu_0/R^\alpha$ (with $\alpha > 0$). Then the leading terms for
large $R$ asymptotic behavior of scalar gravitational degree of freedom
(\ref{eq:phi}) and potential (\ref{eq:pot}) are
\begin{equation}
  \phi \simeq - \frac{\alpha \mu_0}{R^{\alpha+1}}, \hspace{2em}
  V \simeq \text{const} - \frac{(\alpha+1)\mu_0}{3\, R^\alpha}.
\end{equation}
The value of $\phi$ goes to zero in large curvature limit, and the potential
$V$ has power law dependence on $\phi$
\begin{equation}
  V(\phi) \simeq \text{const} - \frac{(\alpha+1)\mu_0}{3\, |\alpha\mu_0|^\gamma}\,\, |\phi|^\gamma,
\end{equation}
with exponent $\gamma$ valued between zero and one
\begin{equation}
  \gamma = \frac{\alpha}{\alpha+1}.
\end{equation}
Thus, the values of both the field and the potential at curvature singularity
are finite for a generic $f(R)$ infrared modification of gravity which
recovers Einstein gravity perturbatively in the large curvature limit. This
means the arguments I made above apply generically, and viability of many
$f(R)$ models in current literature will have to be re-evaluated. At the very
least, the bound (\ref{eq:bound}) will have to be satisfied for the model not
to be ruled out immediately. But even if the estimate for compact objects I
made here looks OK, any infrared-modified $f(R)$ models should be scrutinized
very closely for dangerous curvature singularities that could be present. In a
sense, infrared modification of $f(R)$ gravity forces one to confront the
question of ultraviolet completion of the theory.

Finally, let me comment on how this problem looks like in equivalent
scalar-tensor theory formulation \cite{Whitt:1984pd, Maeda:1988ab}. Conformal
transformation to an Einstein frame with metric $d\hat{s}^2 = f'\,ds^2$ turns
the scalar degree of freedom into a canonically normalized scalar field $\psi$
with potential
\begin{equation}
  \psi = \sqrt{\frac{2}{3}}\, \ln f', \hspace{1em}
  W(\psi) = \frac{1}{2}\, e^{-\frac{4\psi}{\sqrt{6}}}\, (Rf' - f).
\end{equation}
The asymptotics of scalar degree of freedom in Einstein frame are very similar
to the above story: the field $\psi$ goes to zero in large curvature limit,
and the potential has the same unprotected power law asymptotic $W \simeq a -
b\,|\psi|^\gamma$. But where did the singularity go? The answer is subtle:
while the conformal factor itself appears to be regular ($f' \rightarrow 1$),
its second derivatives are not (potential derivative blows up as
$|\psi|^{\gamma-1}$ in equation of motion), which can cause a curvature
singularity in Jordan frame even if Einstein frame metric was regular.

\section*{Acknowledgments}
This work was supported by the Natural Sciences and Engineering Research
Council of Canada under Discovery Grants program. I am grateful to Richard
Battye, Lev Kofman, Levon Pogosian, Ignacy Sawicki, Alessandra Silvestri,
Alexei Starobinsky, and Shinji Tsujikawa for stimulating discussions of
$f(R)$ cosmological models.



\end{document}